\documentclass[twocolumn]{aastex6}

\usepackage{amsmath}
\usepackage{subfigure}

\begin{document}

\title{The Millimeter Continuum Size--Frequency Relationship in the UZ Tau E Disk}
\author{Anjali Tripathi\altaffilmark{1}, 
Sean M.~Andrews\altaffilmark{1}, 
Tilman Birnstiel\altaffilmark{2}, 
Claire J. Chandler\altaffilmark{3}, 
Andrea Isella\altaffilmark{4}, 
Laura M.~P{\'e}rez\altaffilmark{5}, 
R.~J.~Harris\altaffilmark{6}, 
Luca Ricci\altaffilmark{7}, 
David J.~Wilner\altaffilmark{1}, 
John M.~Carpenter\altaffilmark{8},
N.~Calvet\altaffilmark{9},
S.~A.~Corder\altaffilmark{8},
A.~T.~Deller\altaffilmark{10},
C.~P.~Dullemond\altaffilmark{11},
J.~S.~Greaves\altaffilmark{12},
Th.~Henning\altaffilmark{13},
W.~Kwon\altaffilmark{14},
J.~Lazio\altaffilmark{7},
H.~Linz\altaffilmark{13},
L.~Testi\altaffilmark{15,16}
}
\altaffiltext{1}{Harvard-Smithsonian Center for Astrophysics, 60 Garden Street, Cambridge, MA 02138, USA; \url{atripathi@cfa.harvard.edu}}
\altaffiltext{2}{University Observatory, Faculty of Physics, Ludwig-Maximilians-Universit\"at M\"unchen, Scheinerstr.~1, 81679 Munich, Germany}
\altaffiltext{3}{National Radio Astronomy Observatory, P.O. Box O, Socorro, NM 87801, USA}
\altaffiltext{4}{Department of Physics and Astronomy, Rice University, 6100 Main Street, Houston, TX 77005, USA}
\altaffiltext{5}{Departamento de Astronomia, Universidad de Chile, Casilla 36-D, Santiago, Chile}
\altaffiltext{6}{Department of Astronomy, University of Illinois at Urbana-Champaign, 1002 West Green Street, Urbana, IL 61801, USA}
\altaffiltext{7}{Jet Propulsion Laboratory, California Institute of Technology, 4800 Oak Grove Dr., Pasadena, CA 91106, USA}
\altaffiltext{8}{Joint ALMA Observatory, Avenida Alonso de C{\'o}rdova 3107, Vitacura, Santiago, Chile}
\altaffiltext{9}{Department of Astronomy, University of Michigan, 500 Church Street, Ann Arbor, MI 48109, USA}
\altaffiltext{10}{Centre for Astrophysics and Supercomputing, Swinburne University of Technology, PO Box 218, Hawthorn, VIC 3122, Australia}
\altaffiltext{11}{Heidelberg University, Center for Astronomy, Albert Ueberle Str 2, Heidelberg, Germany}
\altaffiltext{12}{School of Physics and Astronomy, Cardiff University, 5 The Parade, Cardiff, CF24 3AA, UK}
\altaffiltext{13}{Max Planck Institut f{\"u}r Astronomie, K{\"o}nigstuhl 17, 69117 Heidelberg, Germany}
\altaffiltext{14}{Korea Astronomy and Space Science Institute, 776
Daedeok-daero, Yuseong-gu, Daejeon 34055, Korea}
\altaffiltext{15}{European Southern Observatory, Karl Schwarzschild Str 2, 85748, Garching, Germany}
\altaffiltext{16}{INAF-Osservatorio Astrofisico di Arcetri, Largo E.~Fermi 5, 50125 Firenze, Italy}

\begin{abstract}
We present high spatial resolution observations of the continuum emission from the young multiple star system UZ Tau at frequencies from 6 to 340\,GHz.  To quantify the spatial variation of dust emission in the UZ Tau E circumbinary disk, the observed interferometric visibilities are modeled with a simple parametric prescription for the radial surface brightnesses at each frequency.  We find evidence that the spectrum steepens with radius in the disk, manifested as a positive correlation between the observing frequency and the radius that encircles a fixed fraction of the emission ($R_{\rm eff} \propto \nu^{0.34\pm0.08}$).  The origins of this size--frequency relation are explored in the context of a theoretical framework for the growth and migration of disk solids.  While that framework can reproduce a similar size--frequency relation, it predicts a steeper spectrum than is observed.  Moreover, it comes closest to matching the data only on timescales much shorter ($\le 1$\,Myr) than the putative UZ Tau age ($\sim$2--3\,Myr).  These discrepancies are the direct consequences of the rapid radial drift rates predicted by models of dust evolution in a smooth gas disk.  One way to mitigate that efficiency problem is to invoke small-scale gas pressure modulations that locally concentrate drifting solids.  If such particle traps reach high continuum optical depths at 30--340\,GHz with a $\sim$30--60\%\ filling fraction in the inner disk ($r \lesssim20$\,au), they can also explain the observed spatial gradient in the UZ Tau E disk spectrum.  
\end{abstract}

\section{Introduction}

One of the fundamental problems in the standard theoretical framework for planet formation occurs at the earliest stages in the planetesimal assembly process.  Small dust grains, incorporated into the protoplanetary disk during the star formation process, grow by collisional agglomeration to modest sizes \citep[$\sim$mm/cm; e.g.,][]{blum08}.  As these grains grow, they begin to dynamically decouple from the gas flow.  In the standard, smooth disk model, a negative radial gas pressure gradient causes the particles to experience aerodynamic drag that results in their inward migration, toward higher gas pressures \citep{adachi76,weidenschilling77,nakagawa86}.  When these {\it radial drift} rates are faster than the collisional growth rate, further growth is effectively halted.  Simulations indicate that this mechanism depletes the outer disk of mm/cm-sized particles on timescales much shorter than disk lifetimes \citep{takeuchi02,takeuchi05,brauer08}, suggesting that planetesimal formation beyond $\sim$10\,au is not possible without invoking some additional, different mechanisms.  

Emission from particles in the mm/cm size range is an important observational tracer of the growth and migration of disk solids.  Theory makes a testable prediction that disks should show a radial segregation of particle sizes, such that the largest particles are concentrated closer to the host star \citep[e.g.,][]{testi14,birnstiel16}.  Fortunately, these particles can be probed directly, with measurements of their thermal continuum emission from (sub)mm to cm wavelengths (frequencies of $\sim$20--700\,GHz; e.g., see \citealt{andrews15}).  Microwave continuum emission from protoplanetary disks is thought to have low optical depths \citep{beckwith90}, meaning the intensity is proportional to the product of the dust opacity, temperature (Planck function), and surface density, $I_\nu \sim \kappa_\nu \, B_\nu(T) \, \Sigma$.  With a rough temperature estimate, the spectral dependence of the continuum emission can be used to determine the shape of the opacity spectrum, $\kappa_\nu$ \citep{beckwith91,ricci10a,ricci10b}, which itself depends on the particle size distribution \citep{miyake93,henning96,dalessio01,draine06}.  

Spatially resolved, multifrequency measurements of the microwave continuum can be employed to test predictions for the spatial variation of the particle size distribution from disk evolution models \citep{isella10}.  A number of studies have used such data in concert with physical prescriptions for disk structure to infer that the microwave spectrum steepens with radius in protoplanetary disks \citep{guilloteau11,banzatti11,perez12,lperez15b,trotta13,menu14,tazzari16}.  In those modeling efforts, such variations were interpreted as spatial gradients in the opacity spectrum, produced by changes in the particle size distribution that are qualitatively consistent with dust evolution theory.  These measurements are definitive, but often cast in the context of physical parameters rather than empirical metrics.  A more direct framing is that the radial brightness profile is found to change as a function of the observing frequency; it becomes relatively more radially extended at higher frequencies.    

In this article, we measure this frequency-dependent variation of the brightness profile in the benchmark disk orbiting UZ Tau E.  
This system is an ideal test case for such measurements  because it is exceptionally bright and spatially extended \citep{tripathi17}, enabling sensitive observations over a wide range of frequencies that are capable of resolving the emission.  UZ Tau E is a close (0.03\,au separation) M1+M4 spectroscopic binary \citep{mathieu96,prato02} located in the nearby Taurus star-forming region.  UZ Tau is a quadruple system, containing another M3+M3 binary pair with a 0\farcs34 ($\sim$48\,au) projected separation (UZ Tau W) that is located 3\farcs8 ($\sim$530\,au) to the west of UZ Tau E \citep[e.g.,][]{simon92}.  The emission from disk material in this system has been studied previously at modest resolution \citep[e.g.,][]{simon92b,jensen96b,guilloteau11,harris12}, demonstrating the presence of three distinct structures: the large circumbinary disk around UZ Tau E and individual disks around each component of UZ Tau W.  

In Section~\ref{sec:data}, we present multifrequency observations from 6 to 340\,GHz of UZ Tau.  In Section~\ref{sec:analysis}, we narrow our focus to the UZ Tau E disk and model its brightness profile at each frequency with a simple parametric prescription.  Section~\ref{sec:results} uses those results to characterize the radial variation of the spectrum in a more empirically-motivated framework.  Section~\ref{sec:discussion} compares the inferred spectral behavior with theoretical models in the context of the evolution of disk solids and recent observations.  Finally, Section~\ref{sec:conclusions} summarizes our conclusions.

\begin{deluxetable*}{llcccccc}[t!]
\tabletypesize{\footnotesize}
\tablewidth{0pt}
\tablecolumns{8}
\tablecaption{Log of UZ Tau Observations\label{table:obslog}}
\tablehead{
\colhead{$\nu$} &
\colhead{Date} &
\colhead{Telescope} &
\colhead{Config.} &
\colhead{Baselines} & 
\colhead{Bandwidth} &
\colhead{Integ.~Time} & 
\colhead{Calibrators} \\
\colhead{[GHz]} &
\colhead{[UTC]} & 
\colhead{} & 
\colhead{} &
\colhead{[m]} & 
\colhead{[GHz]} & 
\colhead{[hours]} &
\colhead{(bandpass, gain, flux)}
}
\startdata
6          & 2011 Jul 23   & VLA   & A             & 680--36000 & 2.0 & 0.3 & 3C\,84, J0431+2037, 3C\,286 \\
30.5/37.5 & 2010 Nov 1    & VLA   & C             & 45--3400   & 1.0 & 1.0 & 3C\,84, J0431+2037, 3C\,286 \\
           & 2011 Mar 19   & VLA   & B             & 240--11000 & 1.0 & 1.0 & 3C\,84, J0431+2037, 3C\,286 \\
           & 2012 Nov 4    & VLA   & A             & 680--36000 & 1.0 & 1.1 & 3C\,84, J0431+2037, 3C\,286 \\
           & 2012 Nov 5    & VLA   & A             & 680--36000 & 1.0 & 1.1 & 3C\,84, J0431+2037, 3C\,286 \\
           & 2012 Nov 6    & VLA   & A             & 680--36000 & 1.0 & 1.1 & 3C\,84, J0431+2037, 3C\,286 \\
           & 2012 Nov 7    & VLA   & A             & 680--36000 & 1.0 & 1.1 & 3C\,84, J0431+2037, 3C\,286 \\
105        & 2009 Dec 17   & CARMA & B             & 82--946    & 2.8 & 3.0 & 3C\,84, J0510+1800, J0510+1800 \\
           & 2010 Mar 24   & CARMA & C             & 26--370    & 2.8 & 2.2 & 3C\,84, 3C\,111, 3C\,111 \\
225        & 2007 Oct 27   & CARMA & C             & 26--370    & 1.9 & 0.7 & 3C\,84, J0530+1331, J0530+1331 \\
           & 2009 Dec 10   & CARMA & B             & 82--946    & 2.8 & 1.1 & 3C\,84, 3C\,111, 3C\,111 \\
           & 2011 Dec 9    & CARMA & A             & 150--1883  & 7.9 & 1.7 & 3C\,84, J0510+1800, J0510+1800 \\
340        & 2010 Feb 18   & SMA   & VEX & 68--509    & 8.0 & 3.2 & 3C\,273, 3C\,111, Vesta \\
           & 2010 Mar 2    & SMA   & VEX & 68--509    & 8.0 & 3.0 & 3C\,273, 3C\,111, Titan \\
           & 2010 Nov 4    & SMA   & COM       & 6--70      & 8.0 & 1.6 & 3C\,273, 3C\,111, Titan \\
           & 2011 Feb 17   & SMA   & EXT      & 16--220    & 8.0 & 1.1 & 3C\,279, 3C\,111, Titan \\
\enddata
\tablecomments{Secondary calibrators (J0510+1800 or 3C\,111) were observed at 340 and 225\,GHz, to check the quality of the phase calibration.  CARMA used regular quasar monitoring bootstrapped to Uranus measurements to set the flux scale.}
\end{deluxetable*}

\begin{deluxetable}{lccc}[b!]
\tablecolumns{4}
\tablecaption{Image Properties\label{table:images}}
\tablehead{
\colhead{$\nu$} &
\multicolumn{2}{c}{Synthesized beam} &
\colhead{RMS noise\tablenotemark{a}} \\
\colhead{[GHz]} &
\colhead{FWHM [arcsec]} & 
\colhead{P.A. [\degr]} & 
\colhead{[mJy beam$^{-1}$]}
}
\startdata
6\phm{space}    & $0.65\times0.36$ & 141 & 0.013 \\
30.5\phm{space} & $0.12\times0.09$ & 120 & 0.006 \\
37.5\phm{space} & $0.10\times0.07$ & 120 & \phm{abcdefg}0.008\phm{abcdefg} \\
105\phm{space}  & $0.93\times0.67$ & 79  & 0.4  \\
225\phm{space}  & $0.19\times0.15$ & 88  & 0.4  \\
340\phm{space}  & \phm{ab}$0.40\times0.29$\phm{ab} & \phm{abc}22\phm{abc}  & 1.3  \\
\enddata
\tablenotetext{a}{Measured in an emission-free region near the image center.}
\end{deluxetable}

\section{Observations and Data Reduction} \label{sec:data}

\subsection{SMA Observations at 340\,GHz}

UZ Tau was observed with multiple configurations of the Submillimeter Array \citep[SMA;][]{ho04} at 340\,GHz (0.88\,mm).  Table~\ref{table:obslog} provides details on the observations.  These SMA data were originally presented by \citet{harris12} and \citet{tripathi17}, but the calibrations for the very extended (VEX) configuration datasets have been updated and improved here.\footnote{A modified baseline solution during this time period was recently derived, which significantly improved the gain calibration and also necessitated a shift in the absolute flux calibration.}  

Observations cycled between the target and nearby quasars on a 10--20 minute cycle.  The data were acquired in good conditions, with precipitable water vapor (PWV) levels $<$2\,mm.  The visibility data from each observation were calibrated independently with standard procedures in the {\tt MIR} package\footnote{\url{https://www.cfa.harvard.edu/~cqi/mircook.html}}.  After correcting for source position shifts and checking for consistency on overlapping baselines, the calibrated visibilities from each observation were spectrally averaged and combined.  The absolute calibration of the visibility amplitudes has a systematic uncertainty of $\sim$15\%.  This composite dataset was Fourier transformed with natural weighting, deconvolved with the {\tt clean} algorithm, and restored with a synthesized beam using {\tt MIRIAD} \citep{sault95}.  A summary of the image properties is provided in Table~\ref{table:images}.

\subsection{CARMA Observations at 105 and 225\,GHz}

The Combined Array for Research in Millimeter-wave Astronomy (CARMA; since de-commissioned) was used to observe UZ Tau at 105\,GHz (2.9\,mm) and 225\,GHz (1.3\,mm): see Table~\ref{table:obslog} for details.  The 225\,GHz observations from 2007 were originally presented by \citet{isella09}.  Observations of UZ Tau were interleaved with visits to nearby quasars on a 5--20\,minute cycle.  The observations were conducted at 225\,GHz with PWV $<$2\,mm ($<$1.5\,mm in the A configuration), and at 105\,GHz with PWV $<2$\,mm in the B configuration and PWV $<$4\,mm in the C configuration.  {\tt MIRIAD} was used for the standard calibration of each individual visibility dataset.  Once calibrated, shifted to account for proper motion, and checked for consistency, the spectrally-averaged visibilities from each observation were combined.  The absolute flux calibration uncertainty is $\sim$15\%\ at both frequencies.  The 225\,GHz visibility data were imaged with natural weighting; the 105\,GHz visibility data were imaged using Briggs weighting with robust = 0.  Table~\ref{table:images} summarizes the synthesized image properties.

\subsection{VLA Observations at 6, 30.5, and 37.5\,GHz}

UZ Tau was observed with the Karl G.~Jansky Very Large Array (VLA) for the ``Disks@EVLA" large program (project code AC982) in the Ka-band, at 30.5\,GHz (9.8\,mm) and 37.5\,GHz (8.0\,mm), and in the C-band at 6\,GHz (5.0\,cm): see Table~\ref{table:obslog}.  The observations alternated between UZ Tau and J0431+2037 on a $\sim$3 and 10 minute cycle for the Ka- and C-bands, respectively.  The data were obtained in good conditions, with low atmospheric optical depths ($\tau \approx 0.03$ at 34\,GHz).  The visibilities for each dataset were calibrated in {\tt CASA} \citep{casa}, using an early version of the scripted VLA calibration pipeline with some additional flagging to remove interference and to ensure an appropriate flux density calibration.\footnote{For more details, see \url{https://science.nrao.edu/facilities/vla/data-processing/pipeline/scripted-pipeline}.}  The systematic uncertainty in the amplitude scale is $\sim$10\%\ at Ka-band and $\sim$5\%\ at C-band.  The individual Ka-band datasets were spectrally averaged into 128\,MHz sub-bands, aligned to ensure that proper motion does not smear the emission, combined, and self-calibrated.  The C-band data were not averaged, to mitigate bandwidth smearing. 

\begin{figure}[t!]
\centering
\includegraphics[width=\linewidth]{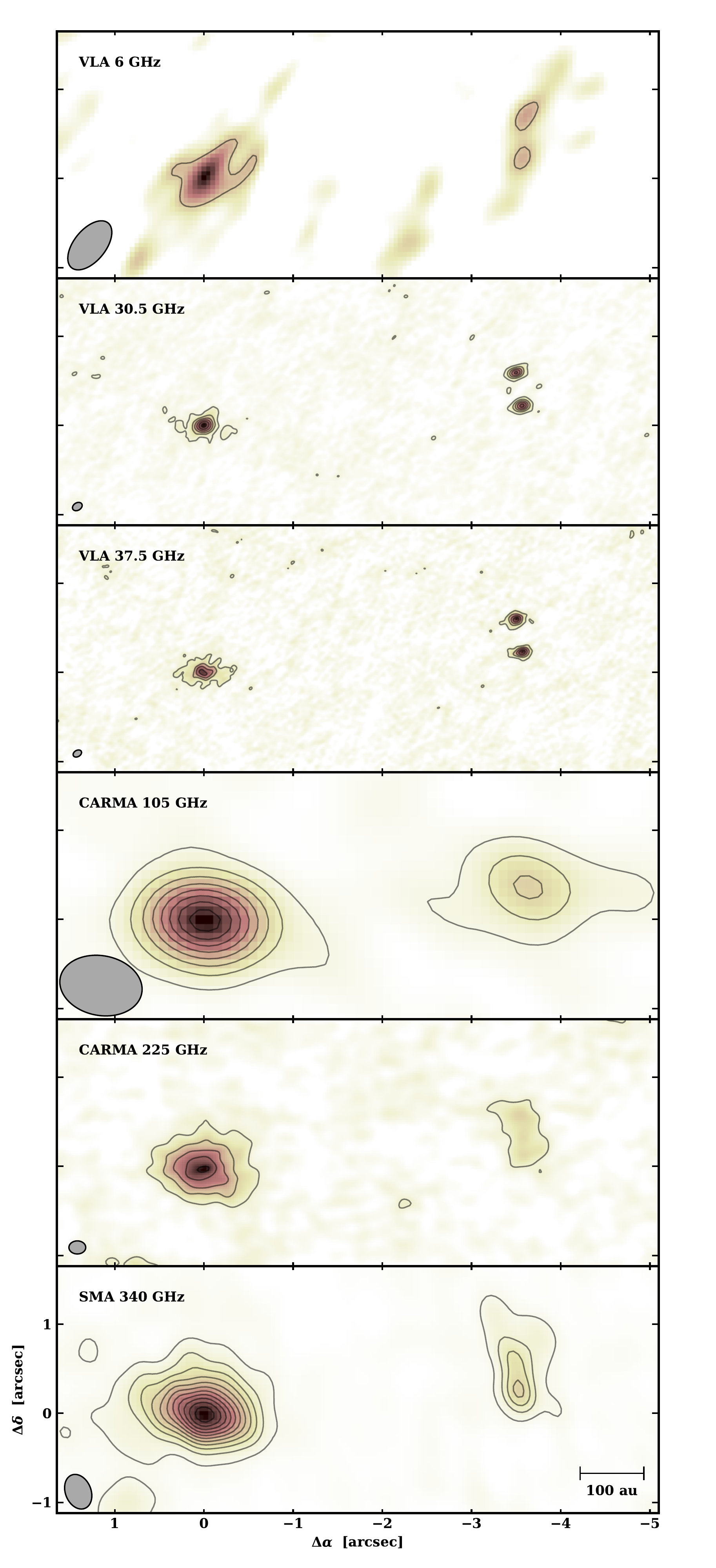}
\figcaption{Maps of the multifrequency continuum emission from UZ Tau.  UZ Tau E is centered at the origin; UZ Tau W is $\sim$3\farcs5 to the west.  Synthesized beam dimensions are shown in the lower left corner of each panel.  Contours are drawn at intervals of 5$\times$ the RMS noise level, starting at 3$\times$ the RMS.  
\label{fig:synthimg}}
\end{figure}

The composite Ka-band dataset was Fourier inverted with natural weighting and then deconvolved using the multi-frequency synthesis version of the {\tt clean} algorithm to account for the flux variation across each sideband.  At C-band, the primary beam is large enough ($\sim$8\arcmin) that any pointing can include some unrelated bright sources.  Failure to {\tt clean} those sources, even if located in the sidelobes of the antenna response pattern, can leave strong residuals near the image center.  Five sources in the UZ Tau field were bright enough to create such imaging artifacts, so we simultaneously deconvolved their outlying fields.  Aside from that caveat, the remainder of the C-band imaging was analogous to the Ka-band.  The image properties in both bands are summarized in Table~\ref{table:images}.

\begin{deluxetable}{lccccc}[t!]
\tablecaption{Estimated Component Spectra\label{table:spectra}}
\tablehead{
\colhead{$\nu$} &
\colhead{$F_\nu$ (E)} &
\colhead{$F_\nu$ (W)} & 
\colhead{$F_\nu$ (Wa)} &
\colhead{$F_\nu$ (Wb)} &
\colhead{Ref.}\\
\colhead{[GHz]} &
\colhead{[mJy]} & 
\colhead{[mJy]} &
\colhead{[mJy]} & 
\colhead{[mJy]} &
\colhead{}
}
\startdata
6    & $0.17\pm0.02$ & $0.18\pm0.03$ & $0.06\pm0.01$ & $0.12\pm0.02$ & 1\\
15 & \multicolumn{2}{c}{$0.48 \pm 0.14$\tablenotemark{a}} & \nodata & \nodata & 2 \\
23 & \multicolumn{2}{c}{$0.77 \pm 0.11$\tablenotemark{a}} & \nodata & \nodata & 2 \\
30.5 & $0.66\pm0.05$ & $0.47\pm0.02$ & $0.24\pm0.01$ & $0.23\pm0.01$ & 1\\
37.5 & $1.08\pm0.07$ & $0.68\pm0.04$ & $0.32\pm0.02$ & $0.36\pm0.03$ & 1\\
43 & $1.8\pm 0.3$ &  & \nodata & \nodata & 2 \\
98 & $14\pm3$ & \nodata & \nodata & \nodata & 3 \\
105  & $22.6\pm0.7$  & $8.5\pm1.0$   & \nodata       & \nodata & 1\\
111 & $22.9 \pm 0.6$ & $6.4 \pm 0.6$ & \nodata & \nodata  & 4 \\
225  & $131\pm6$     & $29\pm3$      & $\sim$16      & $\sim$14 & 1\\
230 & $150 \pm 1$ & $34 \pm 1$ & \nodata & \nodata  & 4 \\
340  & $354\pm13$    & $77\pm9$      & $\sim$23      & $\sim$56 & 1 \\
\enddata
\tablecomments{Flux densities measured from Gaussian fits in the image plane, using the {\tt imfit} task in {\tt CASA}.  These estimates do not include systematic calibration uncertainties.  References (Col.~6): 1 = this work, 2 = \citet{rodmann06}, 3 = \citet{jensen96b}, 4 = \citet{guilloteau11} (note that \citealt{jensen96b} also find consistent 230\,GHz flux densities).}
\tablenotetext{a}{Combined, unresolved flux density for UZ Tau E+W.}
\end{deluxetable}

\subsection{Images and Spectra of UZ Tau E and W}

Figure~\ref{fig:synthimg} shows the synthesized continuum images of the UZ Tau system.  Emission is detected for both the UZ Tau E and W binaries at all six observing frequencies.  The UZ Tau E circumbinary disk is spatially extended at all frequencies aside from 6\,GHz.  Emission contributions from the individual components of the UZ Tau W binary\footnote{For reference, the component to the south is the (optical/infrared) primary, UZ Tau Wa \citep[e.g.,][]{simon92}.} are clearly resolved from one another at 6, 30.5, and 37.5\,GHz but are partially (225, 340\,GHz) or completely (105\,GHz) blended at higher frequencies.  

We used elliptical Gaussian fits (in the image plane) to estimate flux densities for each component in the system and construct the continuum spectra in Table~\ref{table:spectra}.  We will infer the UZ Tau E spectrum again in Section~\ref{sec:results}, from a more rigorous modeling of the visibilities, but those results are consistent with the spectrum in Table~\ref{table:spectra}.   The image-plane fitting is straightforward for UZ Tau E, and for the individual components of UZ Tau W at 6, 30.5, and 37.5\,GHz.  For UZ Tau W at higher frequencies, we attempted simultaneous two-component Gaussian fits with the centers fixed based on the measurements in the Ka-band maps.  At 225 and 340\,GHz, the Wa and Wb component flux densities are rendered considerably uncertain by blending, but the combined W emission is robust (and consistent with aperture photometry).  At 105\,GHz, we only report the combined emission.

\begin{figure}[t!]
\centering
\includegraphics[width=\linewidth]{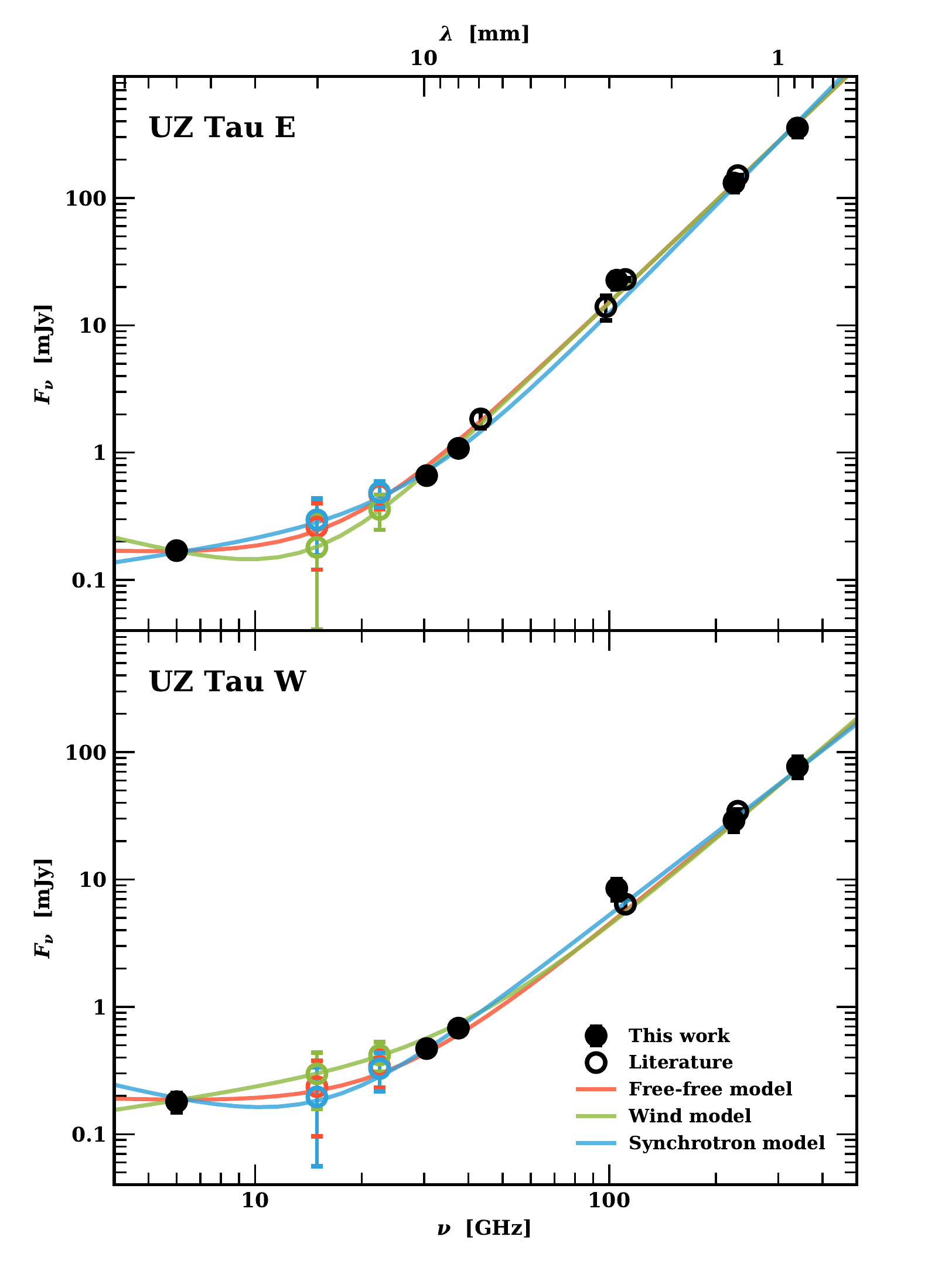}
\figcaption{The spatially integrated radio spectra of UZ Tau E (top) and W (bottom).  Error bars include calibration uncertainties, but are generally smaller than the marker sizes.  The colored curves represent potential model spectra, with the low frequency portion ($<$30\,GHz) constrained by various prescribed decompositions of the UZ Tau E+W combined photometry from \citeauthor{rodmann06}~(colored datapoints), as explained in the text.      
\label{fig:spectrum}}
\end{figure}

The continuum spectra for UZ Tau E and W (Wa + Wb) are shown in Figure~\ref{fig:spectrum}, including some measurements from the literature \citep{jensen96b,rodmann06,guilloteau11}.  Both the E and W spectra can be described with a double power-law, having a steep thermal greybody behavior from dust at higher frequencies ($F_\nu \propto \nu^{\alpha_{\rm d}}$; $\alpha_{\rm d} > 2$) that flattens out or turns over between 6--30\,GHz due to a different, ``non-dust", emission mechanism ($F_\nu \propto \nu^{\alpha_{\rm nd}}$; $\alpha_{\rm nd} \le 0.6$).  

The potential origins of the non-dust emission in UZ Tau E or W are unclear without additional measurements at intermediate frequencies between 6 and 30\,GHz.  Such information is available in the literature, but only as measurements of the {\it combined} (i.e., spatially unresolved) UZ Tau E+W emission \citep{rodmann06}.  Presuming no variability, we can partition that photometry between E and W by assigning an interpretation of the non-dust spectrum from one component.  Figure~\ref{fig:spectrum} illustrates this for three representative interpretations of the W spectrum (i.e., $\alpha_{\rm nd}^{\rm W}$ values).  If the W non-dust spectrum is generated by free-free emission in an optically thin wind ($\alpha_{\rm nd}^{\rm W} = -0.1$; e.g.,  \citealt{mezger67,pascucci12}), then the partitioning of the \citeauthor{rodmann06}~photometry implies that UZ Tau E also has a similar non-dust spectrum ($\alpha_{\rm nd}^{\rm E} \approx -0.1$; {\it red} curves in Figure~\ref{fig:spectrum}).  If the W non-dust contribution is instead from an optically thick and/or structured outflow ($\alpha_{\rm nd}^{\rm W} = 0.4$; e.g., \citealt{reynolds86}), then the E spectrum turns over and has a standard synchrotron spectrum ($\alpha_{\rm nd}^{\rm E} \approx -0.7$, {\it green} curves).  And if W has a synchrotron spectrum ($\alpha_{\rm nd}^{\rm W} = -0.7$), likely produced by stellar activity, then the E non-dust contribution is similar to a structured outflow ($\alpha_{\rm nd}^{\rm E} \approx 0.4$, {\it blue} curves).  In any scenario, the dust contributions have indices, $\alpha_{\rm d}^{\rm E} \approx 2.7$ and $\alpha_{\rm d}^{\rm W} \approx 2.3$, that are typical for disks \citep{ricci10b,ricci10a}.

\subsection{Visibilities for UZ Tau E}

Given the poor resolution of the data compared to the extent of the emission from UZ Tau W, the focus of our analysis will shift exclusively to quantifying the spatial distribution of the multifrequency continuum emitted by the large disk around UZ Tau E.  To derive a UZ Tau E-only suite of visibility datasets, we subtracted the Fourier transforms of the {\tt clean} components (sampled at the observed spatial frequencies) for the UZ Tau W emission (from the maps in Figure~\ref{fig:synthimg}) from the composite datasets.

\section{Analysis of the UZ Tau E Disk} \label{sec:analysis}
\begin{deluxetable*}{ccccccc|c}[t!]
\tablecolumns{10}
\tablecaption{Inferred Surface Brightness Model Parameters\label{table:SBpars}}
\tablehead{
\colhead{$\nu$} &
\colhead{$F_{\nu}$} &
\colhead{$\varrho_t$} &
\colhead{$\log{\alpha}$} &
\colhead{$\beta$} &
\colhead{$\gamma$} &
\colhead{$F_{\rm pt}$} &
\colhead{$\varrho_{\rm eff}$} \\
\colhead{[GHz]} & 
\colhead{[mJy]} & 
\colhead{[arcsec]} & 
\colhead{} & 
\colhead{} & 
\colhead{} & 
\colhead{[mJy]} & 
\colhead{[arcsec]}
}
\startdata 
30.5  & \phd0.71 $^{+ 0.23 }_{- 0.10 }$ & 0.13 $^{+ 0.51 }_{- 0.02 }$ 
      & $p(\alpha); \downarrow$         & $p(\beta); \downarrow$ 
      & 0.95 $^{+ 0.06 }_{- 2.45 }$     & $< 0.19$
      & 0.25 $^{+ 0.26 }_{- 0.04 }$     \\
37.5  & \phd1.14 $^{+ 0.33 }_{- 0.13 }$ & 0.23 $^{+ 0.32 }_{- 0.08 }$ 
      & $p(\alpha); \downarrow$         & $p(\beta); \downarrow$
      & 0.52 $^{+ 0.27 }_{- 2.17 }$     & $<0.24$ 
      & 0.28 $^{+ 0.18 }_{- 0.04 }$     \\
105   & 23.5 $^{+ 2.5 }_{- 0.6 }$       & 0.19 $^{+ 0.64}_{- 0.02 }$
      & $p(\alpha); \downarrow$         & 2.9 $^{+ 4.5 }_{- 0.1 }$
      & $p(\gamma); \uparrow$           & \nodata 
      & 0.41 $^{+ 0.11 }_{- 0.02 }$     \\
225   & 139 $^{+ 11 }_{- 5 }$           & 0.50 $^{+ 0.27 }_{- 0.02 }$
      & 0.46 $^{+ 0.22 }_{- 0.07 }$     & 4.4 $^{+ 3.4 }_{- 0.4 }$
      & 0.31 $^{+ 0.06 }_{- 0.13 }$     & \nodata
      & 0.56 $^{+ 0.05 }_{- 0.02 }$     \\
340   & 368 $^{+ 7\phn}_{- 5 }$         & 0.51 $^{+ 0.04 }_{- 0.02 }$
      & 0.76 $^{+ 0.13 }_{- 0.10 }$     & 3.8 $^{+ 0.4 }_{- 0.2 }$
      & 0.55 $^{+ 0.04 }_{- 0.06 }$     & \nodata
      & 0.62 $^{+ 0.02 }_{- 0.01 }$     \\
\enddata
\tablecomments{The quoted values for each parameter are the peaks of the marginal posterior distributions; uncertainties represent the bounds of the 68.3\%\ confidence interval.  Limits on $F_{\rm pt}$ correspond to the 99.7\%\ confidence boundary.  The notation ``$p(X); \downarrow$'' means the posterior is consistent with the prior (at 95\%\ confidence), but has a marginal preference toward the lower bound (or upper bound, depending on the direction of the arrow).  For clarity, the $F_\nu$ and $F_{\rm pt}$ summaries do not include systematic calibration uncertainties (see Section~\ref{sec:data}); their values would scale with any calibration adjustment, but there would be no effect on the other parameters.  Geometric parameter inferences are described in Section~\ref{sec:analysis}.  Phase center offsets are determined to a precision of $\le$10\,mas.  A visualization of the posterior parameter covariances is presented in the Appendix.}
\end{deluxetable*}

To gain some insight on dust evolution in the UZ Tau E disk, we aim to probe how its microwave continuum spectrum varies with distance from the host stars.  The typical approach for such efforts has been to fit the resolved multifrequency continuum data with a {\it physical} model.  These models parameterize radial variations in the density and particle size distribution (and, thereby, temperature) with reasonable assumptions \citep[e.g.,][]{perez12,perez16,tazzari16}. We opt for a different approach, and instead use empirically-motivated inferences that are more closely related to the {\it observed} continuum morphologies \citep[e.g.,][]{andrews14}.  

The strategy is to use the observed visibilities to constrain the surface brightness profiles at each frequency, $I_\nu(\varrho)$, where $\varrho$ is the radial coordinate projected on the sky \citep[see][]{tripathi17}.  We adopt a parametric brightness profile with the form \citep{lauer95}
\begin{equation}
    I_{\nu}(\varrho) \propto \left( \frac{\varrho}{\varrho_t} \right)^{-\gamma} \left[ 1 + \left (\frac{\varrho}{\varrho_t} \right) ^{\alpha} \right ] ^{(\gamma - \beta)/\alpha},
    \label{eq:nuker} 
\end{equation}
that is characterized by a transition radius ($\varrho_t$), a transition index ($\alpha$), an outer disk index ($\beta$), an inner disk index ($\gamma$), and a normalization (cast with respect to the total flux density, $F_{\nu} \equiv 2\pi \int I_{\nu}(\varrho) \, \varrho \, d\varrho$).  To account for sky projection, we include geometric parameters for the inclination $(i)$, major axis position angle $(\varphi)$, and offsets from the observed phase center.  For the Ka-band models, we also include a point source contribution, centered on the host stars and parameterized by its flux density ($F_{\rm pt}$), to account for any non-dust emission.\footnote{An extrapolation of any reasonable non-dust contribution, like those shown in Figure~\ref{fig:spectrum}, verifies that any such emission can be safely ignored at the higher frequencies.}  We will assess the impact of an alternative non-dust contribution on the resulting inferences in Section~\ref{sec:nondust}.

\begin{figure*}
\centering
\includegraphics[width=\linewidth]{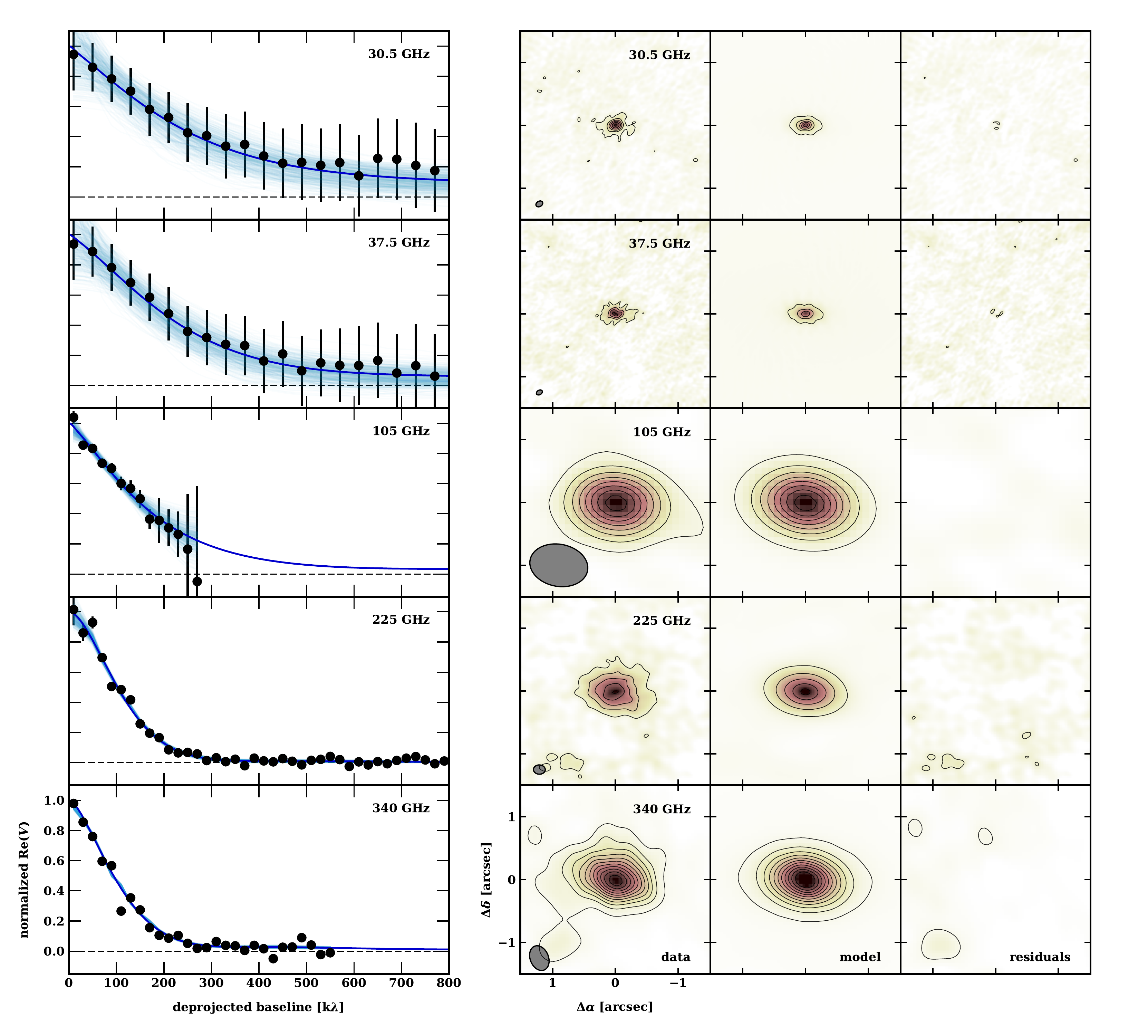}
\figcaption{\textit{Left}: The deprojected and azimuthally averaged real parts of the visibility data (black) with synthetic visibilities constructed from draws from the posterior (cyan).  The mean of those draws, perfectly sampled in the Fourier domain, is overlaid (blue).  All profiles are normalized by the mean flux density of the posterior draws.  \textit{Right}: Synthesized images of the UZ Tau E data (left), model (center), and residuals (right).  The model and residual images are constructed from the mean of the synthetic visibilities derived from the posterior draws (dark blue curves, left panels).  Contours are drawn at intervals of 5$\times$ the RMS noise level, starting at 3$\times$ the RMS.
\label{fig:vis}}
\end{figure*}

For a given set of model parameters, we assume azimuthal symmetry and calculate the two-dimensional Fourier transform of the brightness distribution, sampled at the same spatial frequencies as the observed visibilities.  The data and model are compared with a Gaussian likelihood ($\ln \mathcal{L} \propto -\chi^2/2$).  Prior assumptions on the surface brightness parameters are chosen following \citet{tripathi17}.  For the geometric parameters, we first modeled the well-resolved, high-S/N 340\,GHz data with liberal priors, $p(i) = \sin{i}$ and $p(\varphi) = \mathcal{U}(0$, 180\degr), where $\mathcal{U}$ denotes a uniform distribution over the specified interval.  To ensure consistent projections, we set the priors on $i$ and $\varphi$ at the lower frequencies based on the 340\,GHz marginal posteriors: $p(i) = \mathcal{N}(58\fdg0, 0\fdg7)$ and $p(\varphi) = \mathcal{N}(84\fdg7, 0\fdg7)$, where $\mathcal{N}(\mu, \sigma)$ denotes a normal distribution with the given mean ($\mu$) and standard deviation ($\sigma$).  For the Ka-band point source, we assumed a uniform prior $p(F_{\rm pt}) = \mathcal{U}(0, 1)$\,mJy.  

The posterior distribution of the model parameters conditioned on the data were sampled with the {\tt emcee} package \citep{foreman-mackey13}, which employs the Markov chain Monte Carlo ensemble sampler proposed by \citet{goodman10}.  We assessed convergence as in \citet{tripathi17}; acceptance fractions were in the range 0.2--0.3.  Autocorrelation lengths for all parameters were $\sim 10^2$ steps, implying that (after excising steps for burn-in) we had $\gtrsim10^4$ independent samples of the joint posterior distribution for each frequency. 

\begin{figure}[t!]
\centering
\includegraphics[width=\linewidth]{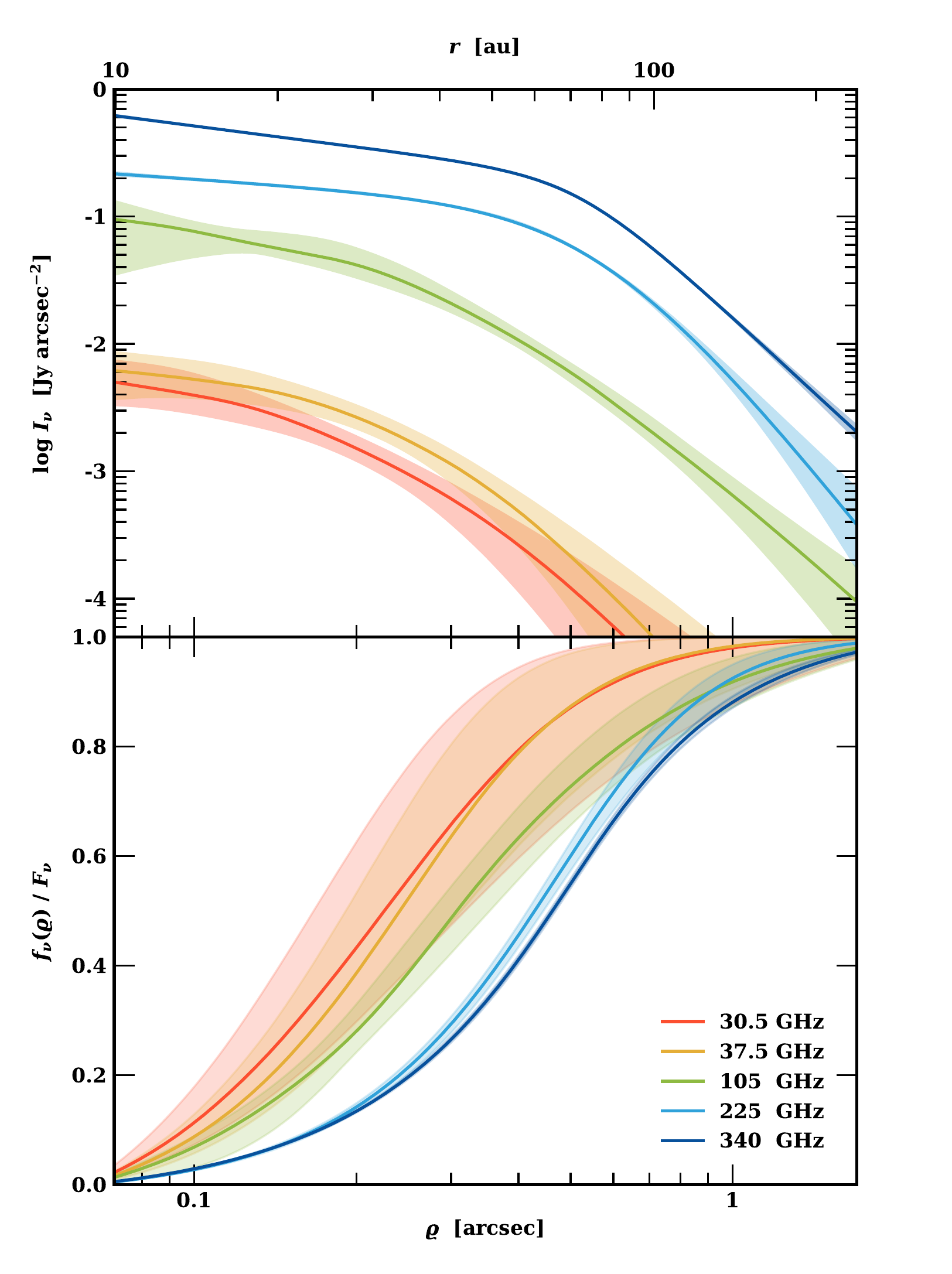}
\figcaption{
\textit{Top:} The 68\%\ confidence intervals for the surface brightness profiles inferred at each frequency.  The median profiles are overlaid as solid curves.    
\textit{Bottom:} The corresponding confidence intervals for the cumulative intensity profiles (see Eq.~\ref{eq:fcum}).  The dashed black horizontal line corresponds to $x = 0.68$, which is used to define the effective size parameter, $\varrho_{\rm eff}$ (see Section~\ref{sec:sizefreq}).
\label{fig:sbprofiles}}
\end{figure}

\section{Results} \label{sec:results}

\subsection{Surface Brightness Inferences \label{sec:fits}}

Table \ref{table:SBpars} summarizes the inferred posterior distributions of the model parameters.  A graphical representation of the posterior covariances is also provided in the Appendix.  Figure~\ref{fig:vis} shows a comparison of the data and model behavior in both the Fourier and image domains.  The left panels show the deprojected, azimuthally-averaged visibilities with corresponding models, constructed from 500 random draws from the posterior.  The model and residual visibility sets made from the average of those draws are used to synthesize representative images.  Overall, the models agree well with the data.

Figure~\ref{fig:sbprofiles} shows the inferred surface brightness profiles and the corresponding cumulative intensity profiles, 
\begin{equation}
f_{\nu}(\varrho) = 2 \pi \int_0^{\varrho} I_{\nu}(\varrho^\prime) \, \varrho^\prime \, d\varrho^\prime,
\label{eq:fcum}
\end{equation}
normalized by $F_\nu$.  Examining all the frequencies together, the different radial extents of the emission can be seen directly from these profiles: the emission is more radially concentrated at lower frequencies.

\subsection{Size--Frequency Relationship \label{sec:sizefreq}}

A useful way to quantify this result is with the size metric introduced by \citet{tripathi17}. We define an {\it effective size} ($\varrho_{\rm eff}$) as the radius that encircles a fixed fraction ($x$) of the total flux density at a given frequency, $f_\nu(\varrho_{\rm eff}) = x F_{\nu}$.  The key advantage of this size metric is that it is largely agnostic of the chosen surface brightness model.  Different model prescriptions yield the same $\varrho_{\rm eff}$ values, provided that they successfully reproduce the data \citep{tripathi17}.  The choice of $x$ is physically arbitrary, although there are practical concerns.  If $x$ is too low, then $\varrho_{\rm eff}$ relies too much on a sub-resolution extrapolation of the brightness profile.  If $x$ is too high, then $\varrho_{\rm eff}$ measurements have inflated uncertainties due to their reliance on faint emission at large $\varrho$.  We adopt $x=0.68$ as a suitable intermediate value to define $\varrho_{\rm eff}$, meant to be crudely comparable to a standard deviation in the approximation of a Gaussian profile. 

\begin{figure}[t!]
\centering
\includegraphics[width=\linewidth]{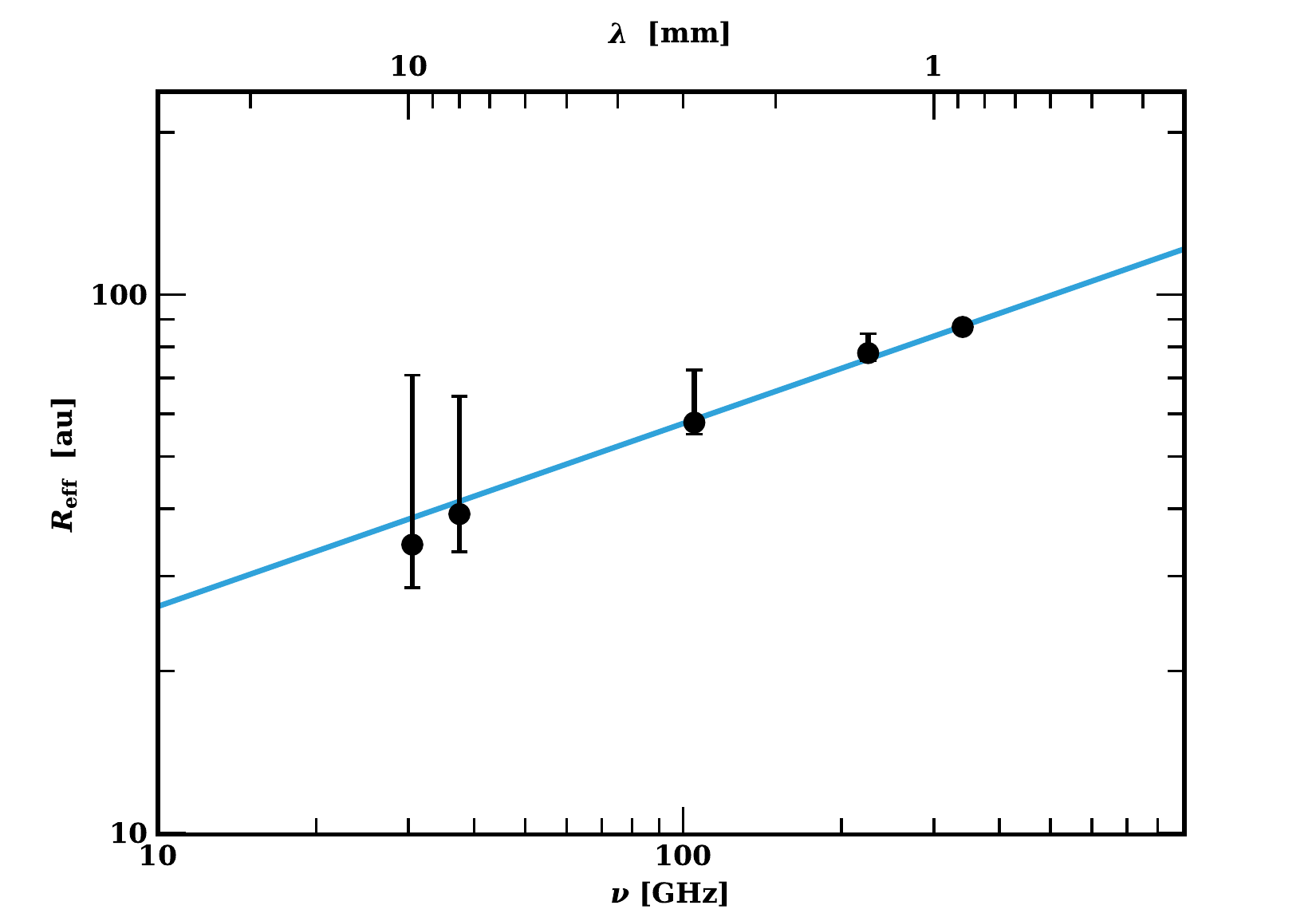}
\figcaption{The inferred size--frequency relationship for the continuum emission from the UZ Tau E disk.  The $R_{\rm eff} \propto \nu^{0.34}$ scaling behavior described in Eq.~\ref{eq:sizefreq} is overlaid (blue).  
\label{fig:sizefreqtrend}}
\end{figure}

Figure~\ref{fig:sizefreqtrend} demonstrates clearly that $\varrho_{\rm eff}$ monotonically increases with the observing frequency (here $R_{\rm eff}$ is an equivalent physical size, assuming a 140\,pc distance to UZ Tau).  A power-law fit to this trend suggests  
\begin{equation}
\log{\left[\frac{R_{\rm eff}}{\rm au}\right]} = (1.1\pm0.2) + (0.34\pm0.08) \log{\left[\frac{\nu}{\rm GHz}\right]},
\label{eq:sizefreq}
\end{equation}
indicating a $>$4\,$\sigma$ deviation from a frequency independent scaling.  As shown in Figure \ref{fig:sbprofiles}, lower S/N data (e.g., at Ka-band) results in larger uncertainties on $\varrho_{\rm eff}$. Nevertheless, the shape (index) of this size--frequency relation is the same (within the uncertainties) for alternative $x$ values in the range 0.5--0.95, but the normalization increases with $x$.  The size--frequency relationship is clear evidence that the continuum spectrum of the UZ Tau E disk steepens with distance from the central stars.

\subsection{Spectral Index Variation \label{sec:alpha_profile}}

\begin{figure}[t!]
\includegraphics[width=\linewidth]{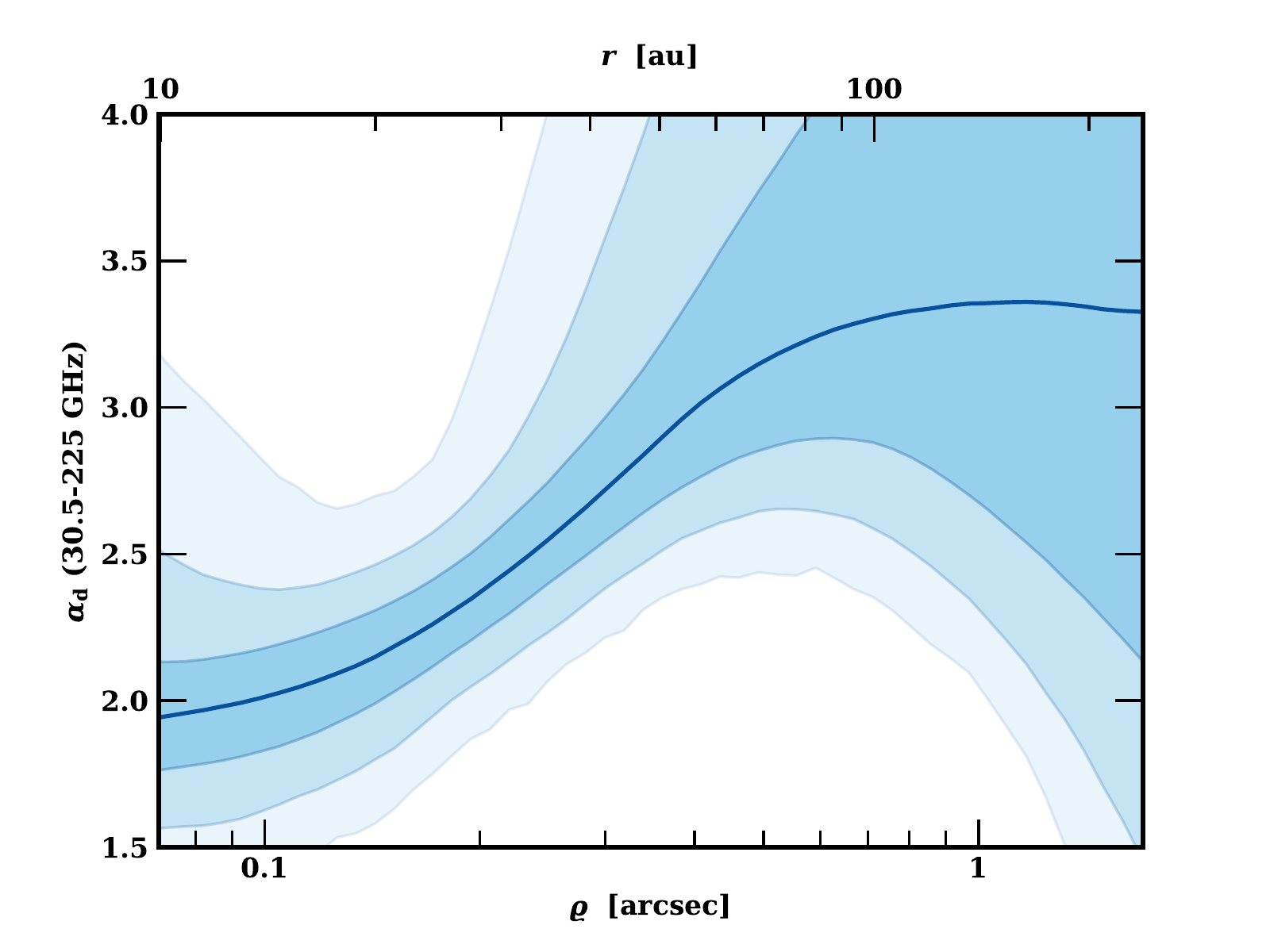}
\figcaption{The continuum spectral index as a function of radius in the UZ Tau E disk, inferred from posterior draws for the 225 and 30.5\,GHz surface brightness models.  From dark to light, the shaded regions correspond to the 68.3, 95.4, and 99.7\%\ confidence intervals.  The median index profile is shown as a dark curve.
\label{fig:alpha}}
\end{figure}

The more traditional way of reaching the same conclusion is to consider physical prescriptions for the radial variation of the opacity spectrum, $\kappa_\nu$ \citep[e.g.,][]{birnstiel10b,perez12,lperez15b, tazzari16}.  Since we are focused on a more empirical interpretation, we instead consider the shape of the emission spectrum.  We can quantify that shape as a ``color" profile, or equivalently a spectral index profile, $\alpha_{\rm d}(r)$.  Spectral index profiles are constructed by taking the ratio of surface brightness profiles at two frequencies, calculated from posterior draws of the model parameters described in Section~\ref{sec:analysis}.   Figure~\ref{fig:alpha} shows the $\alpha_{\rm d}(r)$ profile derived from the 225 and 30.5\,GHz posteriors.  The datasets used to make this inference have roughly matching resolutions (FWHM$\approx$15--20\,au).  The results, shown in Figure \ref{fig:alpha} demonstrate that the spectrum steepens with radius in the disk, with $\alpha_{\rm d}\approx 2$ inside 20\,au and $\alpha_{\rm d} > 3$ outside $\sim$70\,au.  Other frequency pairs show similar behavior (consistent within the uncertainties).   

While an inference of $\alpha_{\rm d}(r)$ is more common (and indeed provides firm evidence of the same behavior), there are good reasons to prefer examining the size--frequency relationship presented in Section~\ref{sec:sizefreq}.  The primary advantage is that the size--frequency scaling simplifies the qualitative interpretation of that spatial variation, since the effective size measurements are robust to our ignorance of the exact form of the brightness profile (or, equivalently the forms of the density, temperature, and opacity profiles).  At the typical resolutions available for this kind of analysis, a variety of profile prescriptions would suitably reproduce the data, and those different forms control the detailed morphology of the inferred $\alpha_{\rm d}(r)$.  On the contrary, any of those prescriptions would produce the same $\varrho_{\rm eff}(\nu)$.  In a sense, the size--frequency relationship is a more compact visualization of the relevant behavior that appropriately acknowledges the limitations in both the data and the model assumptions.

\subsection{Comments on Non-Dust Contributions} \label{sec:nondust}

It is important to understand the nuances of any non-dust contributions in the Ka-band, since those data provide crucial frequency leverage for constraining the spatial variation of the emission spectrum.  In the visibility modeling described above, we have explicitly assumed that any non-dust contribution at these frequencies is point-like (meaning FWHM $\lesssim 0\farcs1$, or 14\,au).  For that assumption, we derived upper limits on any Ka-band non-dust contribution that, together with the flux density at 6\,GHz, imply $\alpha_{\rm nd} < 0.1$.  That spectral index limit is consistent with an optically thin free-free or synchrotron origin (e.g., like the red and green curves, respectively, in Figure~\ref{fig:spectrum}).  We note that \citet{kospal11} speculated that magnetic reconnection events could occur near periastron passage (every 19 days) in the eccentric UZ Tau E binary, and should produce a transient synchrotron spectrum at radio frequencies.  

However, a steeper non-dust contribution to the spectrum could still be accommodated if we relax the assumption of point-like emission.  If the dust contribution to the UZ Tau E spectrum is described by a single power-law, then the combination of the high frequency spectrum and the 6\,GHz flux density suggest a maximum $\alpha_{\rm nd} \approx 0.4$ (steeper indices would significantly over-predict the measured Ka-band flux densities).  Such a spectrum, presumably from a partially thick and/or structured wind \citep[e.g.,][]{panagia75,reynolds86,pascucci12}, would contribute roughly 350\,$\mu$Jy in the Ka-band ($\sim$1/3 of the total flux density).  This is notably more emission than is measured on long baselines, so that emission would need to be spatially resolved.

We do not know a priori what kind of emission distribution would be reasonable in this case, so simplified estimates must suffice.  Presuming a Gaussian distribution, we can place a lower bound of 0\farcs12 (17\,au) on the FWHM of that Ka-band emission.  Following the assumptions of \citet{reynolds86} for such a wind, that limit corresponds to a FWHM $>$0\farcs20 at 6\,GHz, still consistent with the C-band emission being unresolved.  A crude assessment of how this impacts measurements of the Ka-band emission morphology can be made by comparing the Gaussian FWHM required to reproduce the 30.5 and 37.5\,GHz visibilities for cases with and without such wind emission.  We find that assuming a brighter resolved structure, instead of a fainter point-like contribution, would effectively smear out the inferred Ka-band emission distribution from dust by $\sim$10\%.  Without the aid of a well-sampled (and ideally monitored) spectrum from $\sim$6--40\,GHz to directly measure the non-dust emission contribution, this should be considered a systematic (bias) uncertainty on $\varrho_{\rm eff}$ at these frequencies.

\section{Discussion} \label{sec:discussion}

The spatial gradient measured in the microwave continuum spectrum of the UZ Tau E disk can be produced in two different ways, which are not mutually exclusive.  The first scenario considers that the explanation is a corresponding steepening of the dust opacity spectrum with radius.\footnote{Note that a fixed opacity ($\kappa_\nu(r) \approx {\rm constant}$) cannot explain the observations in this scenario (nor if the disk emission were optically thick everywhere).  The decreasing radial temperature profiles in disks would either produce an invariant spectrum across the disk (if all radii are in the Rayleigh-Jeans limit) or one that becomes {\it more shallow} with radius (if the dust at larger radii emits around the peak of the corresponding blackbody curve).}  The shape of the opacity spectrum is largely set by the size distribution of the dust population, such that a more top-heavy size distribution produces a flatter $\kappa_\nu$ \citep{miyake93,henning96,dalessio01,draine06}.  The implication of the spectral variation inferred from the UZ Tau E disk data, and others like it \citep{guilloteau11,perez12,perez16,trotta13,tazzari16}, is that larger particles are concentrated at smaller disk radii, as predicted by dust evolution theory (Section~\ref{sec:dustevol}).  The second scenario posits that the data could also be explained if the emission in the inner disk is confined to high optical depth regions with modest filling factors \citep{ricci12}, where the spectrum should approach $\nu^2$, and the outer disk becomes optically thin (and therefore has a steeper spectrum; Section~\ref{sec:hightau}).  Both possibilities are explored in more detail below.

\subsection{Comparison with Dust Evolution Theory \label{sec:dustevol}}

The first scenario described above hypothesizes that the inferred size--frequency relation is produced naturally by the growth and inward migration (radial drift) of mm/cm-sized ``pebbles" in the disk.  To explore this idea, we compare the observations with a standard theoretical framework for dust evolution in disks based on a coarse grid of simplified models using the code and assumptions presented by \citet{birnstiel12,birnstiel15}.  

This framework presumes an initially homogeneous (in size and dust-to-gas ratio) population of solids embedded in a smooth, viscously evolving gas disk \citep{lyndenbell74,hartmann98}, and then computes the size- and time-dependent surface density profile for the solids.  The gas surface density profile scales like $1/r$ inside a (time-varying) characteristic radius $R_c$, and like $e^{-r}$ at larger radii.  It is normalized by a total (also time-varying) mass $M_d$, and its evolution rate is dictated by a turbulent viscosity parameter $\alpha_t$ (assumed to be constant with time and radius).  The assumed disk temperature profile varies like $r^{-0.5}$, with a normalization that scales with the host star luminosity ($\propto L_\ast^{0.25}$); we impose a fixed minimum temperature of 7\,K.  We use the MIST stellar evolution models \citep{choi16} to compute $L_\ast$  as a function of time, using the sum of the predicted values for a binary host with component masses of 1.0 and 0.3\,$M_\odot$, as inferred dynamically for the UZ Tau E system \citep{simon00,prato02}.  At an age of 1\,Myr, the disk temperature at 10\,au is $\sim$40\,K.  

For any given time and radius, this framework then computes the particle size distribution following the \citet{birnstiel15} prescription, assuming a (spatially and temporally fixed) fragmentation velocity ($u_f$) and ``sticking" probability ($p_s$).  For any particle size, $a$, we compute an opacity spectrum $\kappa_\nu(a)$ based on the assumptions of \citet{ricci10a}, utilizing the optical constants from \citet{weingartner01}, \citet{zubko96}, and \citet{warren84}.  We can then calculate a composite optical depth ($\tau_\nu$) at any frequency, time, and location from the product of the surface densities and opacities, summed over particle size.  From those optical depths and the disk temperatures, we compute radial intensity profiles for the continuum emission, where $I_\nu \sim B_\nu(T) (1 - e^{-\tau_\nu})$.  Those theoretical intensity profiles are used to measure synthetic $F_\nu$ and $\varrho_{\rm eff}$ (as in Section~\ref{sec:analysis}) at the frequencies of interest.  

We consider a coarse grid of evolutionary models for a $M_\ast = 1.3$\,$M_\odot$ host mass \citep{simon00,prato02}, with gas disk structures defined by initial masses $M_d/M_\ast \in [0.01, 0.02, 0.05, 0.1, 0.2]$, initial characteristic radii $R_c \in [20, 40, 100, 200, 300]$\,au, and constant turbulent viscosity coefficients $\alpha_t \in [0.0001, 0.001, 0.01]$ at timesteps from 0.1 to 3\,Myr.  For each disk structure model, we explore constant sticking probabilities $p_s \in [0.1, 0.5, 1.0]$ and fragmentation velocities $u_f \in [1, 5, 10]\,$\,m s$^{-1}$.  For each set of model parameters and each timestep, we compared the model predictions with the data in terms of the emission spectrum (Figure~\ref{fig:spectrum}) and the size--frequency relation (Figure~\ref{fig:sizefreqtrend}).   

\begin{figure}[t!]
\includegraphics[width=\linewidth]{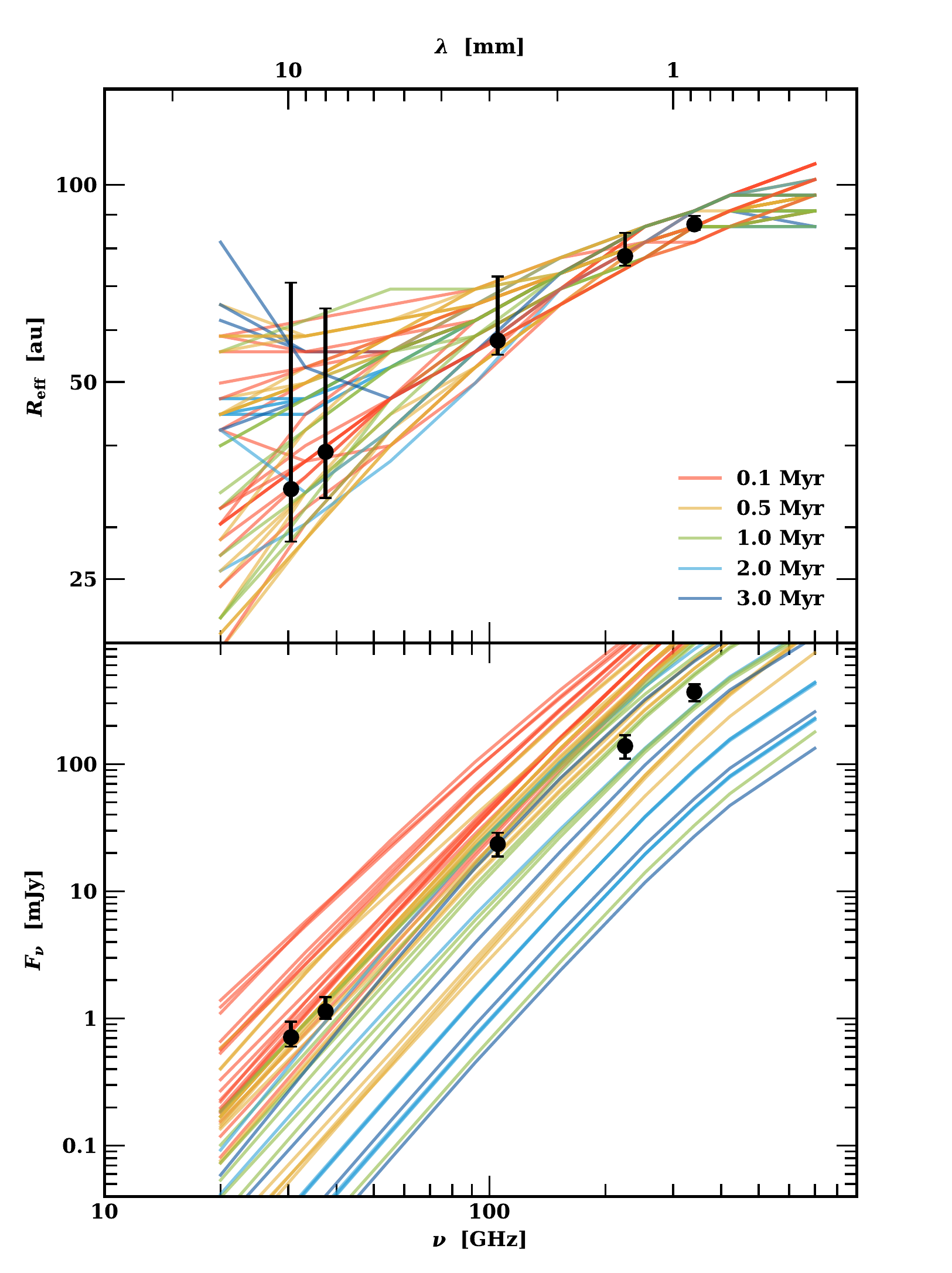}
\figcaption{The dust evolution models that are consistent with the inferred size-frequency relation (top), and their corresponding spectra (bottom), compared with the data.
\label{fig:models}}
\end{figure}

Figure~\ref{fig:models} shows those comparisons for the subset of models that make predictions consistent with the measured size--frequency relation for the UZ Tau E disk (within the 68\%\ confidence intervals).  Models with essentially any disk mass can reproduce this relation, with larger disk masses correlated with smaller characteristic radii (since increasing $M_d$ or $R_c$ increases the $\varrho_{\rm eff}(\nu)$ normalization).  The agreement is best for models with $M_d/M_{\ast} \ge 0.05$ and $40\le R_c\le100$\,au.  Any value of $\alpha_t$, $u_f$, and $p_s$ can reproduce this relation (with appropriate adjustments of other parameters and the evolutionary time), although lower $\alpha_t$ and higher $u_f$ are preferred.  For a given initial gas disk structure and set of microphysics parameters, the evolutionary trend is that the normalization for the size--frequency relation initially increases before dropping over time.  A variety of shapes are possible, but the basic trend is for an overall increase in $\varrho_{\rm eff}$ with $\nu$.  Generally, a lower $p_s$, lower $u_f$, or larger $\alpha_t$ can decrease the overall $\varrho_{\rm eff}(\nu)$ normalization and modestly delay this decay rate, because the growth rates, and thereby the migration rates, of the particles that produce the emission are slowed.  Most of the models that match the size-frequency relation do so at earlier times. 

Figure~\ref{fig:models} also illustrates that all of the models that reproduce the size-frequency relation have a much steeper and/or fainter spectrum than is observed.  While increasing $M_d$ or decreasing $u_f$ can produce more emission, neither is sufficient to reconcile the spectra and $\varrho_{\rm eff}(\nu)$.  The model spectra that are roughly in line with the observations of the UZ Tau E disk occur at early times ($\lesssim$1\,Myr), younger than the nominal $\sim$2--3\,Myr age of UZ Tau E computed from the MIST stellar evolution models and the measured stellar luminosities \citep[e.g.,][]{prato02}.  The normalization of the spectrum decreases with time, such that none of the models at the expected age of UZ Tau E are consistent with the observations.  That said, age constraints on young stars are highly uncertain \citep[e.g.,][]{soderblom14},  especially for accreting close binaries like UZ Tau E \citep[e.g.,][]{stassun14}.  It is possible that a systematic issue could reconcile the apparent timescale discrepancy between the observational constraints and the theoretical predictions.

The same kinds of discrepancies in spectral shape and evolutionary timescales were also noted in previous, similar studies for other disks \citep[e.g.,][]{perez12}.  They are related to the efficiency of particle migration in a gas disk structure with a smooth (i.e., monotonically decreasing) radial pressure profile \citep[e.g.,][]{takeuchi02,takeuchi05,brauer08}.  The radial migration of mm/cm-sized particles in these models is so fast that much of the continuum emission is actually generated by the smaller grains that do not drift.  Relaxing the assumption of monotonic pressure profiles, and instead including substructure in models \citep[e.g.][]{pinilla12b}, will alleviate this timescale discrepancy.

\subsection{Optically Thick Substructures \label{sec:hightau}}

There is an alternative way to explain the observed steepening of the spectrum with radius in the UZ Tau E disk and others like it.  The shallow spectra in the inner disk could be produced by optically thick emission (where $I_\nu \propto B_\nu \sim \nu^2$), and the transition to steeper spectra in the outer disk could mark where the emission becomes optically thin (where $I_\nu \propto \kappa_\nu B_\nu$; since typical particle size distributions have positive microwave opacity slopes, an optically thin spectrum is steeper than $\nu^2$).  This idea has been difficult to reconcile observationally, since the brightness temperatures\footnote{Here we define the brightness temperature using the full Planck equation, $I_\nu = B_\nu(T_b)$, since the Rayleigh-Jeans approximation is not applicable for the temperatures and frequencies of interest.} ($T_b$) measured in the inner disk are typically lower than the expected (beam-averaged) dust temperatures ($T_d$).  However, \citet{ricci12} rightly pointed out that this scenario is still feasible if the optically thick emission is concentrated on size scales smaller than the resolution.     

For the specific case of the UZ Tau E disk, the spectrum varies like $\nu^2$ out to a radius of $\sim$20\,au, before transitioning to a much steeper $\nu^{3.5}$ by $\sim$100\,au (see Figure~\ref{fig:alpha}).  We measure roughly the same peak $T_b \approx 10$--15\,K inside a radius of 20\,au from 30 to 340\,GHz.  Assuming the temperature prescription adopted in Section~\ref{sec:dustevol} (for ages of 1--3\,Myr), the (beam-averaged) $T_d$ in that same region would be 25--35\,K.  To reconcile the expected $T_d$ and measured $T_b$ in this scenario, the filling factor for optically thick emission needs to be $\sim$0.3--0.6.          

These characteristics are not unique to the UZ Tau E disk; recent observations provide some precedent that clearly associates such behavior with small-scale, optically thick substructures in protoplanetary disks.  High resolution observations of the disks around HL Tau \citep{brogan15} and TW Hya \citep{andrews16} show that their inner regions exhibit fine-scaled concentric rings of emission with similar (areal) filling factors and high continuum optical depths, with a transition to more optically thin emission at larger radii \citep{tsukagoshi16,carrasco16,liu17,huang17}.  Moreover, those disks have similar peak $T_b$ for the same frequencies at comparable resolutions to those measured here for the UZ Tau E disk \citep{kwon11,kwon15,carrasco16,andrews12,menu14}.

Given the apparent prevalence of such small-scale substructures in disks \citep[e.g.,][]{zhang16,cieza16,isella16} and the plausibility that the associated emission has higher optical depths than is typically assumed, this option for explaining the measured size--frequency relationships seems quite promising.  Such features might be markers of {\it local} maxima in the gas pressure distribution that preferentially concentrate migrating particles \citep[e.g.,][]{whipple72,klahr97,pinilla12b}.  The possibility that such particle traps mitigate the classical problem of fast radial drift rates makes this a compelling and natural solution to the apparent discrepancies noted in comparisons between data and dust evolution models (e.g., Section~\ref{sec:dustevol}).  The fundamental question of plausibility is whether or not these disks exhibit such small-scale continuum modulations when observed at higher angular resolution.

\section{Conclusions} \label{sec:conclusions}

We have presented and analyzed high resolution observations of the continuum emission from the UZ Tau system at frequencies of 6, 30.5, 37.5, 105, 225, and 340\,GHz.  For the large disk around the UZ Tau E binary, we modeled the visibility data at each frequency from 30.5--340\,GHz independently with a simple prescription for the surface brightness distribution, and then used those results to help characterize the spatial variation of the microwave continuum spectrum across the disk.  A positive correlation between the inferred {\it size} of the emission and its observing frequency is found, providing clear evidence that the spectrum steepens with radius.  

We have considered whether standard models for the evolution of solids embedded in a {\it smooth} gas disk could explain these observations.  While these models predict qualitatively similar behavior, they evolve too quickly compared to the inferred age of UZ Tau E.  Instead, we suggest that the most likely origin for the measured size--frequency relation is the presence of small-scale (i.e., unresolved), optically thick substructures in the inner disk.  Our expectation is that higher resolution continuum observations of the UZ Tau E disk (e.g., using long ALMA baselines) will reveal such substructures.  If that prediction is confirmed, the observed substructure morphology could be used to modify the assumption of a smooth gas disk in the dust evolution models \citep[e.g., following][]{pinilla12b} and help explore how particle traps influence the growth and migration of solids.

\acknowledgments We thank an anonymous reviewer for useful suggestions to improve the manuscript.  We are grateful to Jane Huang for helpful discussions, and Mark Gurwell and Shelbi Schimpf for deriving an improved baseline calibration for the SMA data.  S.A. acknowledges support for this work from NASA Origins of Solar Systems grant NNX12AJ04G.  T.B. acknowledges funding from the European Research Council (ERC) under the European Union's Horizon 2020 research and innovation programme under grant agreement No 714769.  J.M.C. acknowledges support from the National Aeronautics and Space Administration under grant No. 15XRP15\_20140 issued through the Exoplanets Research Program.  Part of this research (by J.L.) was carried out at the Jet Propulsion Laboratory, California Institute of Technology, under a contract with the National Aeronautics and Space Administration.  This work (by L.T.) was partly supported by the Italian Ministero dell\'\,Istruzione, Universit\`a e Ricerca through the grant Progetti Premiali 2012 -- iALMA (CUP C52I13000140001), and by the Deutsche Forschungs-gemeinschaft (DFG, German Research Foundation) - Ref no. FOR 2634/1 TE 1024/1-1.
CARMA  development and  operations were supported by the National Science Foundation  under  a cooperative agreement and by the CARMA partner universities.  The Submillimeter Array is a joint project between the Smithsonian Astrophysical Observatory and the Academia Sinica Institute of Astronomy and Astrophysics and is funded by the Smithsonian Institution and the Academia Sinica.  The  VLA is run by the National  Radio Astronomy Observatory, a facility of the  National Science Foundation operated under cooperative agreement by Associated Universities, Inc.

\facilities{CARMA, SMA, VLA}
\software{{\tt  Astropy} \citep{astropy}, {\tt CASA} \citep{casa}, {\tt corner} \citep{corner}, {\tt emcee} \citep{foreman-mackey13}, {\tt Matplotlib} \citep{matplotlib}, {\tt Miriad} \citep{sault95},  {\tt MIR} (\url{https://www.cfa.harvard.edu/~cqi/mircook.html})}

\appendix
\section{Posterior Distributions}
\setcounter{figure}{0}  
\begin{figure}[h]
\centering
\centering
\includegraphics[width=.8\linewidth]{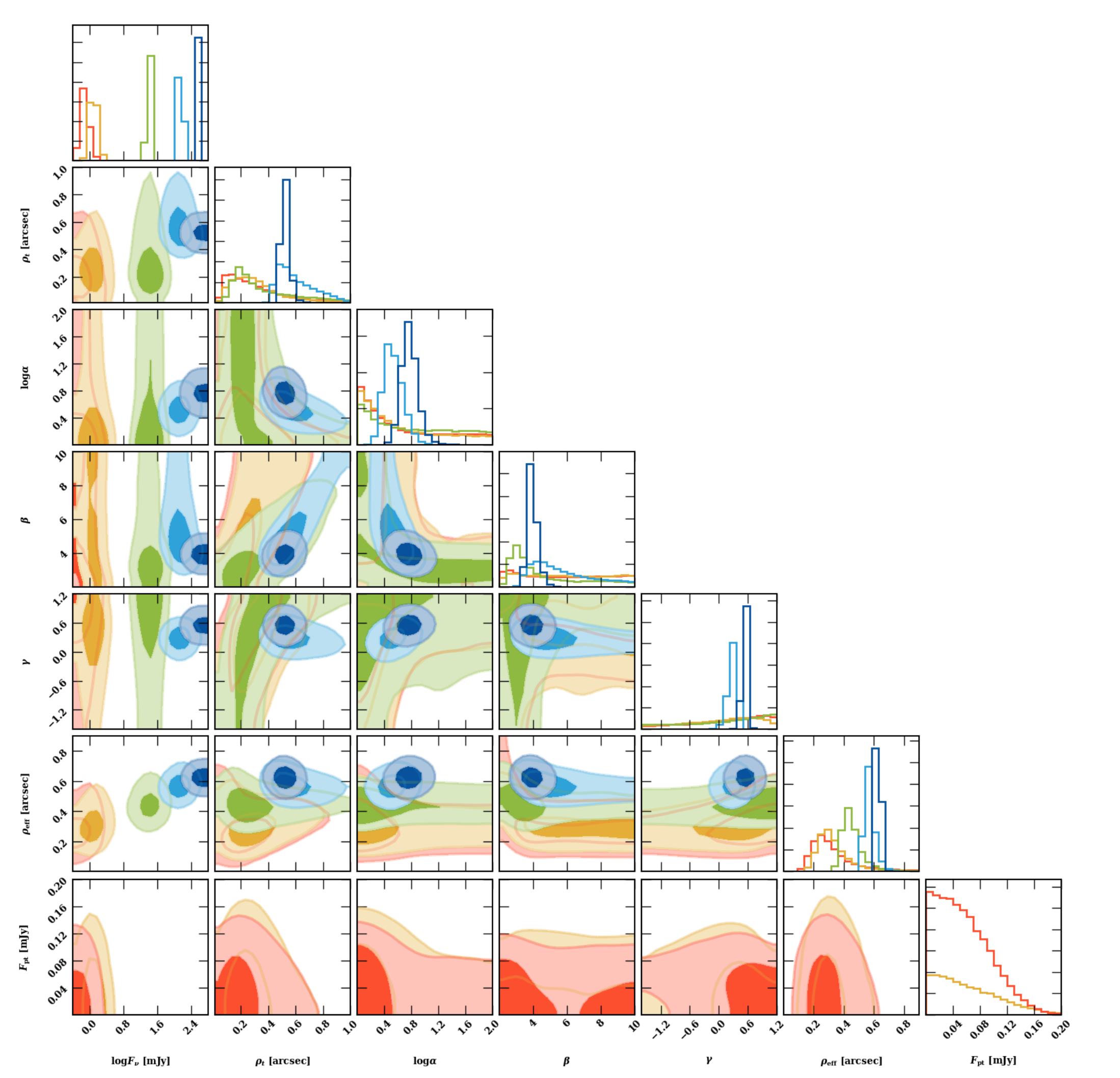}
\figcaption{Covariances between the surface brightness model parameters and the effective size (colors as in Fig. \ref{fig:sbprofiles}).  Only the Ka band models include the $F_{\rm pt}$ parameter.
\label{fig:covar}}
\end{figure}

\clearpage
%\bibliography{references}

\end{document}